\documentclass[twocolumn,preprintnumbers,amsmath,amssymb,aps,pra,longbibliography]{revtex4-1}
\usepackage{amsmath,amssymb,graphicx,epstopdf,bm,soul,upgreek, natbib}
\usepackage{graphicx}
\usepackage{color}
\definecolor{OliveGreen}{rgb}{0,0.6,0}
\usepackage{dcolumn}
\usepackage{bm}
\usepackage{amssymb,float}
\usepackage{cancel}
\usepackage{ulem}
\usepackage[T1]{fontenc}
\usepackage[colorlinks,bookmarks=false,citecolor=blue,linkcolor=blue,urlcolor=blue]{hyperref}

\begin{document}

\title{From the topological spin-Hall effect to the non-Hermitian skin effect in an elliptical micropillar chain}%

\author{Subhaskar Mandal}
\email{subhaska001@e.ntu.edu.sg}
\author{Rimi Banerjee}\email{rimi001@e.ntu.edu.sg}
\author{Timothy C.H. Liew}\email{tchliew@gmail.com}

\affiliation{Division of Physics and Applied Physics, School of Physical and Mathematical Sciences, Nanyang Technological University, Singapore 637371, Singapore}

\begin{abstract}
 The topological spin-Hall effect causes different spins to propagate in opposite directions based on Hermitian physics. The non-Hermitian skin effect causes the localization of a large number of modes of a system at its edges.  Here we propose a system based on exciton-polariton elliptical micropillars hosting both the effects. The polarization splitting of the elliptical micropillars  gives rise to the topological spin-Hall effect in a one dimensional lattice.  When a circularly polarized external incoherent laser is used to imbalance effective decay rates of the different spin polarizations, the system transits to a non-Hermitian regime showing the skin effect. These effects have implications for robust polariton transport as well as the deterministic formation of multiply charged vortices and persistent currents.
\end{abstract}

\maketitle


\section*{Introduction}
The field of topological insulators (TIs) started with the discovery of the quantum Hall effect \cite{PRL.45.494.1980} followed by the Haldane model \cite{PRL.61.2015.1988}, where electrons propagate robustly along the edges of a finite 2D sample. Later, Kane and Mele came up with the spin dependent version of the Haldane model, where they predicted that electrons with different spins would propagate robustly in opposite directions \cite{PRL.95.226801.2005}. This is known as the topological spin-Hall effect (TSHE), which is expected to play an important role in the field of spintronics \cite{NatNanotech.9.794.2014}.

\begin{figure}[t]
\centering
\includegraphics[width=0.47\textwidth]{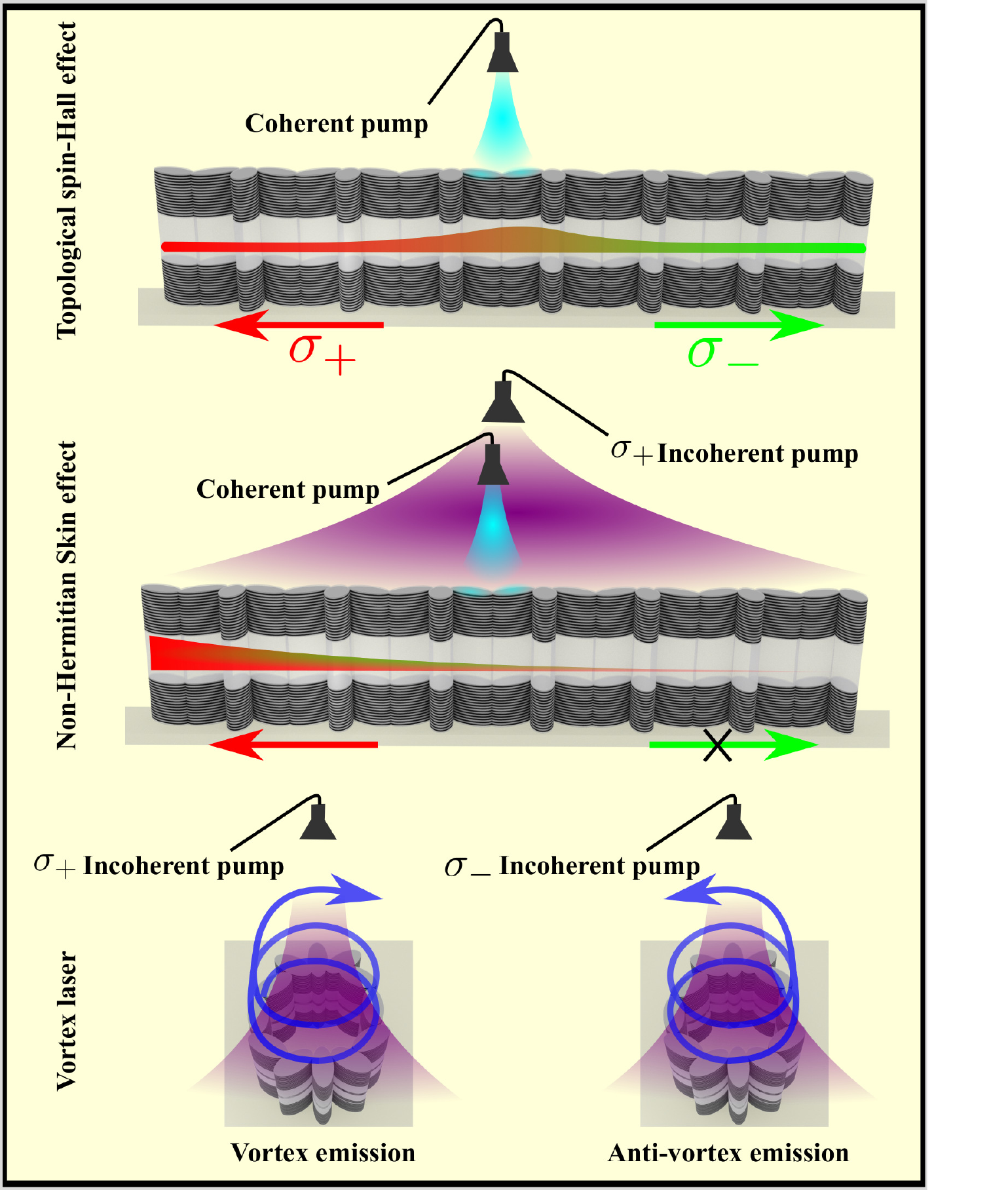}
\caption{Schematic of an exciton-polariton elliptical micropillar chain. Proper orientations of the elliptical micropillars naturally give rise to the spin-Hall effect. If the polaritons having opposite circular polarizations $\left(\sigma_\pm\right)$ have unequal decay rates, which can be arranged by subjecting the whole  chain to a circularly polarized uniform incoherent pump, the non-Hermitian skin effect appears. In a circular chain, polariton condensation in a vortex state can be achieved, which can act as a vortex laser.}
\label{Fig1}
\end{figure}

In the last decade, the concept of TIs has not only been extended from fermions to different bosonic systems such as photonics \cite{PRL.100.013905.2008,Nature.461.772.2009,NatPhoton.8.821.2014,RevModPhys.91.015006.2019,Nature.565.622.2019} and acoustics \cite{PRL.114.114301.2015}, but higher order TIs have also been demonstrated in those systems \cite{PRB.98.205147.2018,NatCom.11.3768.2020,NatMater.18.108.2019,PRL.122.244301.2019,NatPhys.NonLinear.2021,Light.Sci.Appl.10.164.2021}. In a photonic system the role of spins can be played by the photonic circular polarizations $\left(\sigma_\pm\right)$. However, due to the presence of the splitting in energy between the transverse-electric (TE) and transverse-magnetic (TM) modes (which couples the $\sigma_\pm$ polarizations), realizing  the TSHE for photons is far from trivial and often needs extremely careful fabrication in order to reduce the TE-TM splitting \cite{NatMater.12.233.2013}. Instead of the photonic polarization, if the role of spin is played by an external degree of freedom such as the clockwise and counter-clockwise propagation directions in ring resonator lattices \cite{NatPhys.7.907.2011,PRL.110.203904.2013,Science.367.59.2020} or valley degree of freedom \cite{NJP.18.025012.2016,PRL.120.063902.2018,NatPhoton.14.446.2020}, then realizing the TSHE in photonic systems is more readily achieved.

Recently, non-Hermitian topological phases, especially the non-Hermitian skin effect (NHSE), have been an intense area of research. Typically in a finite 1D lattice if the coupling between the adjacent sites is made different  in amplitude  in the forward and backward directions (non-reciprocal coupling), then all the modes of the system get localized at the edges \cite{PRL.121.086803.2018,PRB.97.121401R.2018,PRB.99.201103R.2019,PRL.124.056802.2020,NatCom.11.5491.2020,PRL.121.073901.2018,NatPhys.16.747.2020,PRL.125.226402.2020,PRL.125.186802.2020,arXiv:2104.08844}. This is unlike the topological Hermitian systems, where the number of edge modes residing within the bulk band gaps is determined by the bulk-edge correspondence \cite{RevModPhys.82.3045.2010}. In the NHSE any injected signal, independent of its excitation position, always propagates in one direction while the propagation in the opposite direction is strongly suppressed \cite{PRL.121.073901.2018,Science.368.311.2020}. Such an effect would be useful in suppressing feedback in optical circuits, which is a necessity for their function \cite{Science.230.138.1985}.   However, achieving the non-reciprocal coupling between two sites in an optical system requires amplification of the coupling along a particular direction between a pair of sites \cite{Science.368.311.2020} or requires manipulation of gain and decay in each unit cell \cite{AnnalenderPhysik,PRL.125.033603.2020,srep13376}.

Exciton-polaritons are quasiparticles arising due to the strong coupling of cavity photons and quantum well excitons \cite{RevModPhys.85.299.2013}. Their finite lifetime, sensitivity to external fields, strong non-linearity inherited from excitons, and spin polarizations  make them ideal for studying both the Hermitian and non-Hermitian topological phases. Polariton Chern insulator analogues relying on Zeeman splitting and TE-TM splitting are well-established \cite{PRX.5.031001.2015,PRB.91.161413.2015,PRL.114.116401.2015,PRB.94.115437.2016,PRB.97.081103R.2018,PRB.100.235444.2019,PRAppl.12.064028.2019,OptLett.45.5311.2020,PRAppl.12.054058.2019,Nature.562.552.2018}. Nonlinear topological polaritons have also been an intense area of research \cite{Optica.3.1228.2016,PRL.119.253904.2017,PRL.124.063901.2020,OptLett.45.4710.2020}, where in some cases the nonlinearity alone induces topological behavior \cite{PRB.93.020502R.2016,PRB.96.115453.2017,PRB.99.115423.2019,Arxiv.2102.13285.2021}. The inherent non-Hermitian nature of the polaritons has lead to the realization of exceptional points \cite{Nature.526.554.2015,arxiv.2012.06133.2021}, non-Hermitian topological phases \cite{NatCom.11.4431.2020,arxiv.2012.06133.2021}, and other related theoretical predictions \cite{PRA.95.023836.2017,PRR.2.022051.2020,PRB.104.195301.2021}. The spin-orbit coupling due to the TE-TM splitting inside the microcavity has played a vital role in obtaining nontrivial spin related effects, such as the optical spin Hall effect \cite{NatPhys.3.628.2007,PRL.109.036404.2012,LightSciAppl.7.74.2018}, meron polarization textures \cite{Optica.8.255.2021}, and nontrivial bands with spin orbit interaction Hamiltonians \cite{PRB.98.155428.2018,arxiv.1912.09684.2019,Science.366.727.2019,NatPhoton.2020,Optica.8.255.2021,NatCom.12.689.2021,Optica.7.455.2020,PRB.103.L081406.2021}.

Here, we propose a system based on an exciton-polariton elliptical micropillar chain, where both the TSHE and NHSE can be realized in a single set up [see Fig.~\ref{Fig1}]. Proper orientation of the otherwise identical elliptical micropillars, along with the naturally present polarization splitting due to their shape anisotropy gives rise to a band structure where the positive and negative momentum states have opposite circular polarizations $\left(\sigma_\pm\right)$, i.e., the TSHE. If the effective decay rates of the $\sigma_\pm$ polaritons are made unequal by subjecting the chain to a circularly polarized spatially uniform incoherent pump,  non-Hermitian physics takes over. The behavior of the system changes drastically and the modes of the system get localized at one end of the chain, giving rise to the NHSE \cite{PRL.121.086803.2018}, which is further confirmed by calculation of the relevant topological invariant (the winding number of the complex energies in the Brillouin zone). The key advantage of our scheme is that the robust polariton propagation here is in a one dimensional lattice, which is more compact for relevant information transport applications.   This system shows large non-reciprocity over topologically trivial waveguides. By arranging the chain into a closed ring, the spinor eigenstates separate into multiply charged vortices and deterministic persistent currents appear naturally [see Fig.\ref{Fig1}].
\begin{figure}[t]
\centering
\includegraphics[width=0.47\textwidth]{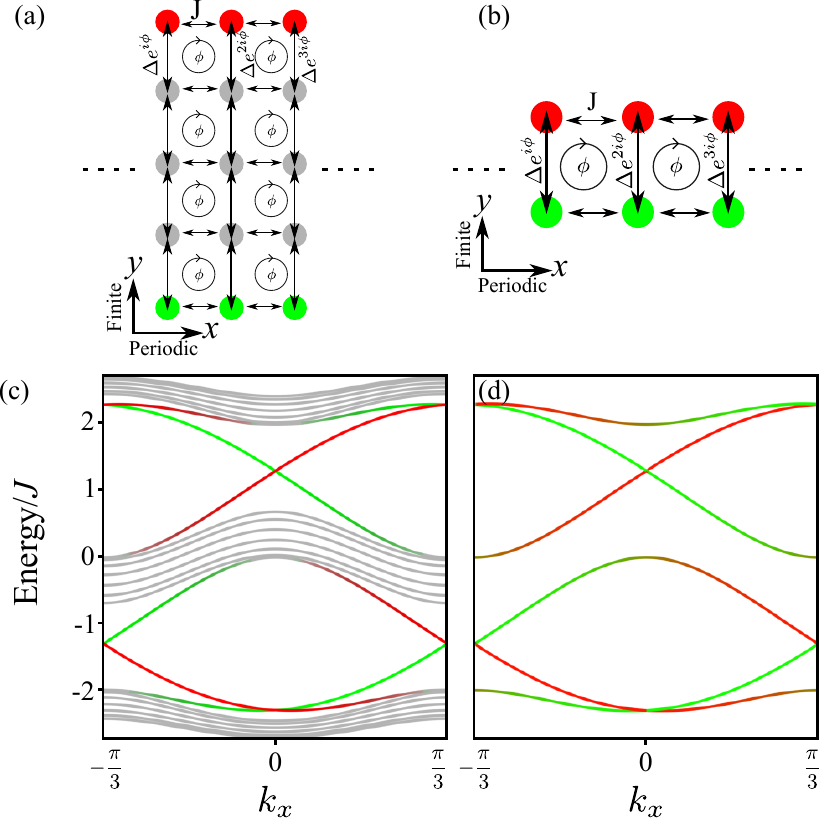}
\caption{(a) The unit cell of a square lattice with real couplings along the $x$ direction and complex couplings along the $y$ direction such that each plaquette has total accumulated phase $\phi=2\pi/3$. (b)The same lattice as in (a), but having only two sites along the $y$ direction. (c)-(d) The band structure of the system presented in (a)-(b), respectively. The bulk bands are shown in grey, whereas red and green being the edge states located at the opposite edges. Parameters: $\Delta/J=1$.}
\label{Fig2}
\end{figure}

\section*{Topological spin-Hall effect} We first consider a square lattice, which is periodic along the $x$ direction and has $N_y$ sites along the $y$ direction [see Fig.~\ref{Fig2}(a)]. The coupling between the nearest sites along the $x$ direction is taken as $J$, while the coupling between the nearest sites along the $y$ direction is taken as complex $\Delta e^{in_x\phi}$, where $n_x$ is the site number along the $x$ direction. The total phase in each plaquette becomes $\phi$ ($-\phi$) while hopping counter-clockwise (clockwise). Any nonzero phase $\phi$ introduces an effective magnetic field, which breaks the time reversal (TR) symmetry in the system. The Hamiltonian corresponding to the system is  

\begin{align}\label{Eq1}
\hat{H}=&\sum_{n_x,n_y} -\left[J \hat{a}^{\dagger}_{n_x,n_y}\hat{a}_{n_x+1,n_y}+\Delta e^{in_x\phi}\hat{a}^{\dagger}_{n_x,n_y}\hat{a}_{n_x,n_y+1}\right]\nonumber\\
&+ \text{h.c.}
\end{align}
Here $n_y$ is the site index along the $y$ direction, $\hat{a}^{\dagger}_{n_x,n_y}\left(\hat{a}_{n_x,n_y} \right)$ is the particle creation (annihilation) operator at site $\left(n_x,n_y\right)$, and h.c. represents the Hermitian conjugate. The Hamiltonian (\ref{Eq1}) gives rise to the Hofstadter butterfly if the eigenenergies of the system are plotted against $\phi$ \cite{NatPhys.7.907.2011,PRB.98.075412.2018}, where for $\phi=2\pi/q$ each Bloch band splits into $q$ sub-bands. In Fig.~\ref{Fig2}(a) the unit cell of the system is shown for $\phi=2\pi/3$. The band structure of the system is shown in Fig.~\ref{Fig2}(c), where the bulk bands are shown in grey, while red and green represent the edge states located at the two edges. As expected, there are three bulk bands, connected by topological edge states. 

Next, we reduce the width of the lattice and keep only two sites along the $y$ direction [see Fig.~\ref{Fig2}(b)], giving the band structure shown in Fig.~\ref{Fig2}(d). Remarkably the band structure is the same as the one in Fig.~\ref{Fig2}(c), however all the bulk modes disappear leaving behind only the edge modes \cite{PhysRevA.89.023619.2014}. 

\begin{figure}[t]
\centering
\includegraphics[width=0.47\textwidth]{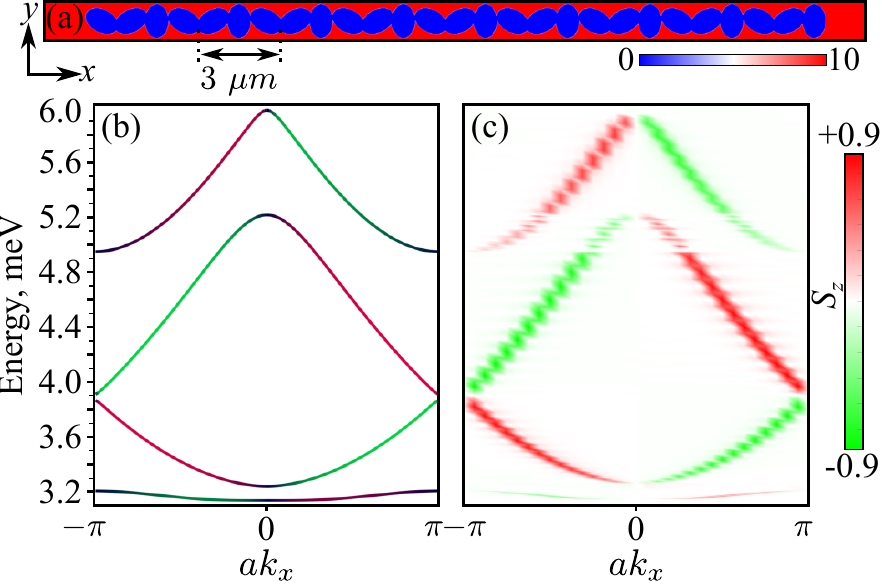}
\caption{(a) A chain of elliptical micropillars, where three pillars oriented by $\pi/3$ with respect to each other form an unit cell. (b)-(c) Band structure of the system under consideration using Bloch theorem (infinite chain) and by Fourier transforming the eigenstates of a finite chain.}
\label{Fig3}
\end{figure}

Considering the red (green) sites as $\sigma_+$ ($\sigma_-$)  polariton spin states, the system shown in Fig.~\ref{Fig2}(b) can be mapped to a chain of elliptical micropillars (see Fig.~\ref{Fig3}(a)), where $\Delta$ is the polarization splitting inside the micropillars due to shape anisotropy \cite{PRB.100.115305.2019} and $\phi$ represents the orientation angle. With this interpretation the system becomes TR symmetric. The states having positive and negative momentum have opposite circular polarizations, corresponding to the TSHE.

The complex hopping between the $\sigma_\pm$ states inside an elliptical micropillar can be shown using a basis transformation, where an orientation angle  $\theta$ induces $2\theta$ phase in the hopping term (see supporting information).  The dynamics of the polaritons in a micropillar chain is described by the driven-dissipative Gross-Pitaevskii equation coupled with an exciton reservoir

\begin{align}
i\hbar\frac{\partial\psi_{\sigma_{\pm}}}{\partial t}&=\left[-\frac{\hbar^2\nabla^2}{2m}+V(x,y)-i\hbar\frac{\gamma}{2}\right]\psi_{\sigma_{\pm}}\notag\\
&+V_T(x,y,\theta)\psi_{\sigma_\mp}+\left[\alpha_1\left|\psi_{\sigma_\pm}\right|^2+\alpha_2 \left|\psi_{\sigma_\mp}\right|^2\right]\psi_{\sigma_\pm}\notag\\
&+\left[g_r+i\hbar\frac{R}{2}\right]n_{\sigma_{\pm}}\psi_{\sigma_{\pm}}+F_{\sigma_{\pm}}(x,y)e^{-i\omega_p t},\label{Eq4}\\
\frac{\partial n_{\sigma_{\pm}}}{\partial t}&=-\left[\gamma_r+R|\psi_{\sigma_{\pm}}|^2\right]n_{\sigma_{\pm}}+\xi\left[n_{\sigma_\mp}-n_{\sigma_\pm}\right]+P_{\sigma_{\pm}}\label{Eq4B}.
\end{align}
Here $\psi_{\sigma_\pm}$ represents the wave functions of the $\sigma_\pm$ polariton spin states, $\nabla^2$ represents the Laplacian, $m=7.5\times10^{-5}m_e$ is the polariton mass with $m_e$  the free electron mass. $V$ represents the potential corresponding to the chain of elliptical micropillars shown in Fig.~\ref{Fig3}(a) with potential depth 10 meV. To realize $2\pi/3$ phase in each plaquette (see Fig.~\ref{Fig2}(b)), any two neighbouring elliptical micropillars need to be oriented by an angle $\pi/3$ with respect to each other. To be consistent with experiment we choose the micropillars having semi-major axis 1.3 $\mu$m and semi-minor axis $0.85~\mu$m \cite{PRB.100.115305.2019}, which gives a three micropillar unit cell with length $a = 3~\mu$m. $V_T$ represents the linear polarization splitting around 1 meV inside the pillars \cite{PRB.100.115305.2019}, which has the same spatial profile as $V$ but multiplied by a factor $e^{\pm2i\theta}$ corresponding to the orientation of each micropillar.   $\alpha_1~(\alpha_2)$ is the polariton-polariton interaction constant between same (opposite) spins.  $n_{\sigma_\pm}$, $g_r$, and $R$ represent the exciton density in the reservoir, interaction of the polaritons with the reservoir, and the condensation rate, respectively. $\gamma$ and $\gamma_r$ are the polariton and reservoir decay, respectively. $\xi$ represents the spin relaxation process in the reservoir, which plays a crucial role for circularly polarized incoherent excitation \cite{PRL.109.016404.2012}. $P_{\sigma_\pm}$ is a uniform incoherent pump and $F_{\sigma_{\pm}}$ is a coherent pump having frequency $\omega_p$.

In Fig.~\ref{Fig3}(b) we remove the terms related to pump, decay and non-linearity ($P_{\sigma_\pm}=\gamma=\gamma_r=R=g_r=\xi=F_{\sigma_\pm}=\alpha_1=\alpha_2=0$)  and calculate the band structure using the Bloch theorem. In Fig.~\ref{Fig3}(c) the same band structure is obtained by Fourier transforming the eigenstates of a finite chain, which corresponds to the photoluminescence spectrum in the experiments. The colorbar corresponds to the  degree of circular polarization, which is defined as $S_z=\left(|\psi_{\sigma_+}|^2-|\psi_{\sigma_-}|^2\right)/\left(|\psi_{\sigma_+}|^2+|\psi_{\sigma_-}|^2\right)$. 

\begin{figure*}[t]
\centering
\includegraphics[width=\textwidth]{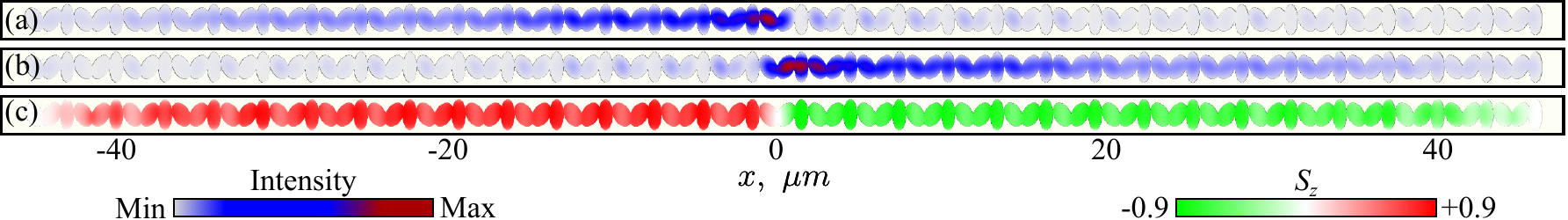}
\caption{Topological spin-Hall effect: (a)-(c) $|\psi_{\sigma_+}|^2$, $|\psi_{\sigma_-}|^2$, and $S_z$, respectively,  at $t=60$ ps under a linearly polarized coherent pump positioned at the middle ($x=0~\mu$m) of the chain. Parameters: $\gamma=0.05$ ps$^{-1}$, $\alpha_1=1~\mu$eV$\mu$m$^2$, $\alpha_2=-0.1\alpha_1$, peak value of $F_{\sigma_\pm}=0.4$ meV$\mu$m$^{-1}$, $\hbar\omega_p=4$ meV, and $g_r=R=\gamma_r=\xi=P_{\sigma_\pm}=0$.}
\label{Fig4}
\end{figure*}

To illustrate the TSHE, we inject the polaritons using a tightly localized linearly polarized Gaussian shaped coherent continuous pump positioned at the middle of the chain. Here the reservoir dynamics is neglected.  While modelling the polariton dynamics under coherent excitation only, the dynamics of the reservoir is typically neglected \cite{PRL.93.166401.2004}. It is a very good approximation since the exciton reservoir, which lies above the injected polariton energy is unlikely to be excited. This results in very small or no population of the excitons in the reservoir, which hardly impacts the coherently injected polaritons. However, in presence of an incoherent excitation, which we will consider in the next section, this approximation is no longer valid and one needs to account for the gain and potential shift caused by the presence of the reservoir. The full width at half maximum of the pump is around 2.1 $\mu$m, which excites a wide range of momentum corresponding to a particular energy. In Figs.~\ref{Fig4}(a) and \ref{Fig4}(b) the dynamics of the $\sigma_\pm$ polaritons are shown, where they propagate in opposite directions showing the TSHE. Such non-trivial propagation of polariton spins exists for an energy range around 1.5 meV (3.3 - 4.8 meV; above this the propagation direction for $\sigma_\pm$ polaritons is reversed). This is around 15 times larger than the non-trivial bandgap of the topological polaritons observed in \cite{Nature.562.552.2018}. Fig.~\ref{Fig4}(c) shows the degree of circular polarization of the propagating polaritons (and Movie  1 shows the full dynamics).  

\section*{Non-Hermitian skin effect} 
\begin{figure*}[t]
\centering
\includegraphics[width=1\textwidth]{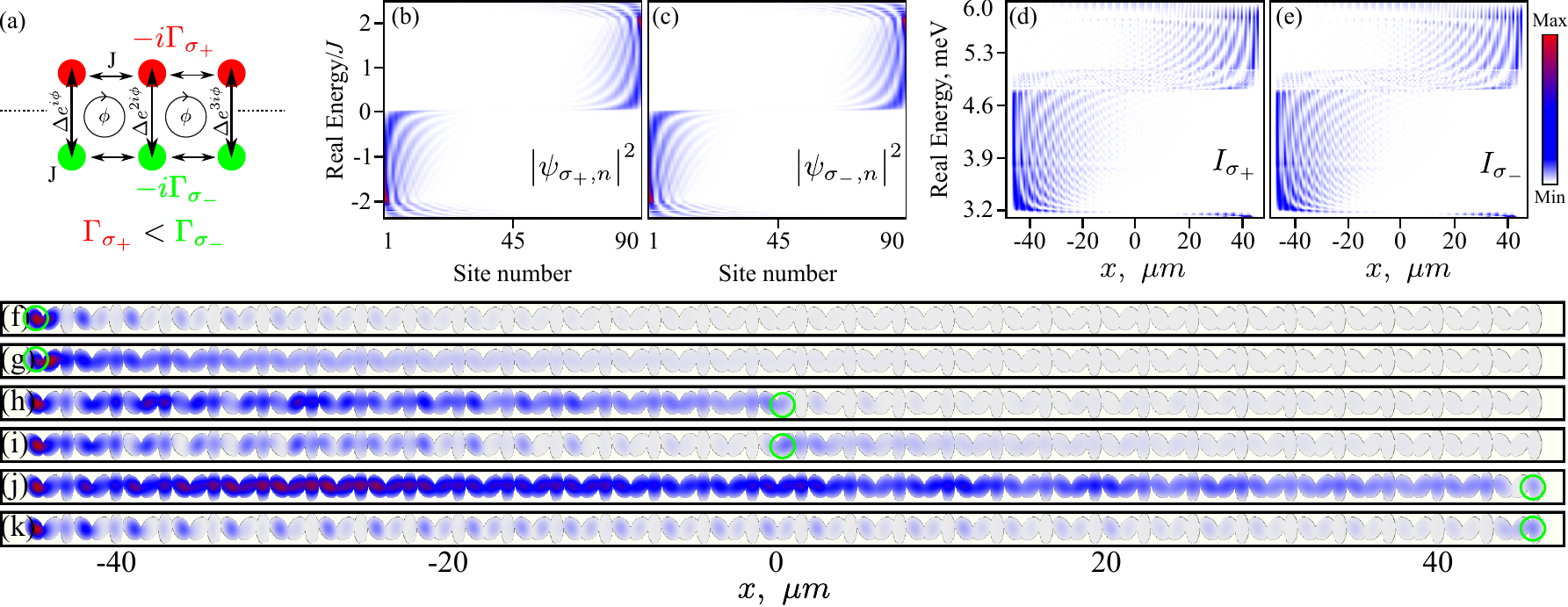}
\caption{(a) Scheme of the non-Hermitian skin effect where the decay rates of $\sigma_\pm$ modes are different.  (b)-(c) Mode localization using the tight binding model in Eq.~(\ref{Eq5}). Parameters:  $\Delta/J=1$, $\left(\Gamma_{\sigma_-}-\Gamma_{\sigma_+}\right)/J=0.3$. (d)-(e) Plot of $I_{\sigma_\pm}(x)$ for each eigenmode calculated from Eq.~(\ref{Eq4}), which shows the mode localization. Polariton propagation at $t=200$ ps under a linearly polarized coherent pump when the pump is positioned at the left end (f)-(g), middle (h)-(i) and the right end of the chain (j)-(k). (f), (h), (j) correspond to $|\psi_{\sigma_+}|^2$ and (g) ,(i), (k) correspond to $|\psi_{\sigma_-}|^2$.  Parameters: $\gamma=0.5$ ps $^{-1}$, $\gamma_r=1.5\gamma$, $g_r=2\alpha_1$, $R=10$ ps$^{-1}\mu$m$^2$, $\xi=1.6$ ps$^{-1}$, $P_{\sigma_+}=0.07$ ps$^{-1}\mu$m$^{-2}$, and $P_{\sigma_-}=0$. The green circles represent the excitation position. All other parameters are the same as those in Fig.~\ref{Fig4}.}
\label{Fig5}
\end{figure*}

So far we have discussed Hermitian physics, with the non-Hermitian decay only contributing a factor $e^{-\gamma t}$ in intensity.  However, if we consider different decay rates for the $\sigma_\pm$ modes, the situation changes drastically. The Hamiltonian in Eq.~(\ref{Eq1}) becomes:

\begin{align}\label{Eq5}
\hat{H}=&-J\sum_{n_x,\sigma_\pm} \left[ \hat{a}^{\dagger}_{n_x,\sigma_{\pm}}\hat{a}_{n_x+1,\sigma_{\pm}}+\hat{a}^{\dagger}_{n_x+1,\sigma_{\pm}}\hat{a}_{n_x,\sigma_{\pm}}\right]\nonumber\\
&+\sum_{n_x,\sigma_\pm}\left[-\Delta e^{\pm in_x\phi}\hat{a}^{\dagger}_{n_x,\sigma_{\pm}}\hat{a}_{n_x,\sigma_{\mp}}-i\Gamma_{\sigma_\pm}\hat{a}^{\dagger}_{n_x,\sigma_\pm}\hat{a}_{n_x,\sigma_\pm}\right].
\end{align}
Here $\Gamma_{\sigma_\pm}$ are the effective decay rates of the $\sigma_\pm$ modes. We choose $\Gamma_{\sigma_+}<\Gamma_{\sigma_-}$ and diagonalize the Hamiltonian having 90 sites along the $x$ axis [see Fig.~\ref{Fig5}(a)]. The spatial profiles of all the modes corresponding to both $\sigma_{\pm}$ spins are plotted in Figs.~\ref{Fig5}(b) and \ref{Fig5}(c). The modes having real energy $<0$ ($>0$) are localized at the left (right) end of the lattice.  This can be confirmed as the NHSE by calculating the winding of the complex energy in the Brillouin zone, which is a topological invariant (see sec.~\ref{SEC4}).  The localization of the polaritons can be understood in the following way: $\sigma_\pm$ polaritons propagate in opposite directions.  When $\sigma_+$ polaritons hit the end of the chain, they couple to counter-propagating $\sigma_-$ modes. However, since $\Gamma_{\sigma_+}<\Gamma_{\sigma_-}$, the $\sigma_-$ modes decay  (much faster compared to $\sigma_+$ modes) before propagating a significant distance. Due to this a large population of the polaritons builds up at the end of the chain giving rise to the NHSE. The group velocity of the modes with smallest decay (here $\sigma_+$) decides the localization end of the bands. If one considers the case where $\Gamma_{\sigma_+}>\Gamma_{\sigma_-}$, then $\sigma_-$ polarized modes become those with smallest decay, which have  group velocity opposite to the $\sigma_+$ modes. Consequently, the localization is in the opposite end in this case (see Movie  2).

In the real physical system, we would expect that $\sigma_+$ and $\sigma_-$ polarized polaritons would decay with the same rate. However, if we introduce a spatially uniform $\sigma_+$ polarized incoherent pump, it can be expected to provide a gain that compensates the decay for $\sigma_+$ component such that effective decay rates for $\sigma_+$ and $\sigma_-$ components will be different. We can model this explicitly using Eqs.~(\ref{Eq4}) and (\ref{Eq4B}). Ideally, we want to provide gain only for $\sigma_+$ polaritons, keeping $n_{\sigma_-}=0$. However, due to the spin relaxation, the reservoir is  partially spin polarized. This effect is taken into account by the $\xi$ term in Eq.~(\ref{Eq4B}). We note that the degree of circular polarization (DOCP) of the photoluminescence is shown to be around 18\% for a bare quantum well \cite{NatPhoton.13.283.2019} and we set $\xi=1.6$ ps$^{-1}$ to give a matching value of the reservoir spin polarization. Next, we diagonalize the Hamiltonian corresponding to Eq.~(\ref{Eq4}) by taking the steady state of the reservoir and define, $I_{\sigma_\pm}(x)=\int|\psi^E_{\sigma_\pm}|^2~dy$, for each eigenstate $\psi^E_{\sigma_\pm}$ and plot it as a function of the real energies and the $x$ axis in order to see the localization of the eigenstates corresponding to a finite chain [see Figs.~\ref{Fig5}(d) and \ref{Fig5}(e)]. The states are localized at the edges, consistent with the tight binding model. To check the polariton dynamics, we set the incoherent pump near but below the condensation threshold and inject the polaritons using the same coherent pump as in the TSHE case. In Figs.~\ref{Fig5}(f) and \ref{Fig5}(g) the coherent pump is positioned at the left end of the chain, which shows that the polaritons do not propagate from left to right even after 200 ps.  Next the coherent pump is positioned at the middle of the chain, where polaritons accumulate at the left end. The drastic difference in the polariton dynamics from the TSHE can be seen by comparing Figs.~\ref{Fig5}(h) and \ref{Fig5}(i) with Figs.~\ref{Fig4}(a) and \ref{Fig4}(b). At initial times $\sigma_-$ polaritons try to propagate  towards the right but quickly decay and at larger times when the steady state is reached the total polariton population is at the left end  (see Movie  3). This is understandable as the NHSE arises due to the combination of TSHE and different decay rates of the $\sigma_\pm$ modes. The accumulation of the $\sigma_-$ modes at the left end of the chain mainly arises due to the spin flipping of the $\sigma_+$ modes. Finally in Figs.~\ref{Fig5}(j) and \ref{Fig5}(k) the coherent pump is positioned at the right end of the chain, where $\sigma_\pm$ polaritons propagate from right end to left end and accumulate at the left end. This shows that the one-way propagation of the polaritons is independent of the position of injection. The localization plotted in Figs.~\ref{Fig5}(d) and \ref{Fig5}(e) corresponds to the eigenstates of the system, which are localized at one end of the chain. To observe the injected polaritons accumulate at one end of the chain we must wait till the steady state is reached. However, at transient times, we can already observe that the left and right propagating polaritons have different intensities (see the pulsed excitation in the supporting information). 

Although, we have used coherent excitation for injecting polaritons into the system, it is possible to set the strength of the circularly polarized spatially uniform incoherent pump above the condensation threshold (see the supporting information). For such a case the condensation spontaneously forms at the end of the chain. 

\section*{Topological invariant of the skin modes}\label{SEC4}
\begin{figure}[t]
\centering
\includegraphics[width=0.45\textwidth]{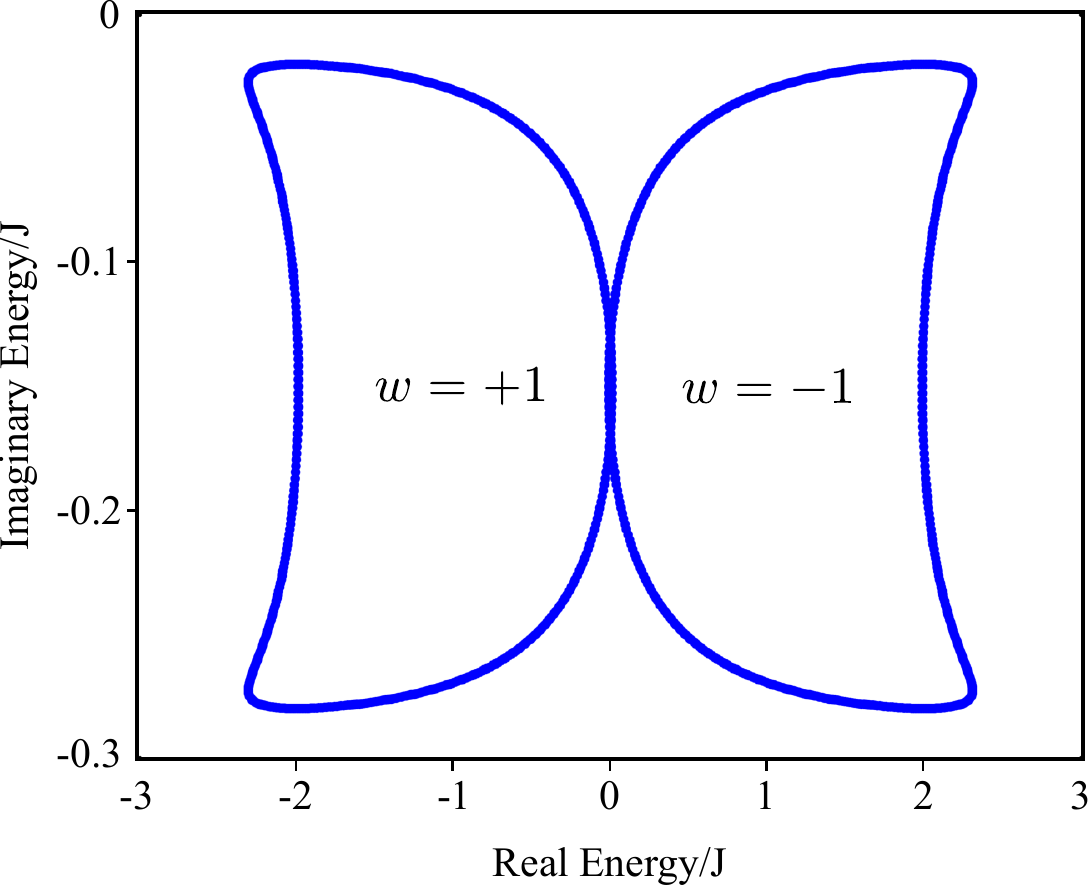}
\caption{Plot of real vs imaginary energy, which forms a loop having non zero winding number. All the parameters are the same as those in Figs.~\ref{Fig5}(b) and \ref{Fig5}(c), but using the periodic boundary condition.}
\label{FigWinding}
\end{figure}
The eigenenergies of a Hermitian system are always real. Consequently, the bandgap of a Hermitian system can be defined as the difference in energies between the top and bottom of  two consecutive bands. However, the energy bands corresponding to a non-Hermitian system are generally complex. As a result, the definition of bandgap used in a Hermitian system is no longer valid. The bandgap in a non-Hermitian system can be characterized in two ways: point gap and line gap \cite{PRX.8.031079.2018}. A non-Hermitian Hamiltonian is said to have a point gap with respect to a reference point $E_R$ in the complex plane (real  vs imaginary energy plane) if the complex bands do not cross the reference point $E_R$. Similarly for a line gap the complex bands should not cross the reference line. Non-Hermitian Hamiltonians having a line gap can be transformed through a unitary transformation to Hermitian Hamiltonians \cite{PRX.8.031079.2018}. As a result the topological properties of these non-Hermitian Hamiltonians can be linked to the Hermitian topology of the transformed Hermitian Hamiltonians. On the other hand, there is no Hermitian counter part of the non-Hermitian Hamiltonian with a point gap. The non-Hermitian skin effect is a consequence of non-trivial topology of a point gapped non-Hermitian system.

The relevant topological invariant for the non-Hermitian skin effect is the winding number $w$, which can be defined with respect to a reference energy $E_R$  as \cite{PRX.8.031079.2018}

\begin{align} \label{Eq1_100}
w=\sum_{n=1}^N\oint_{\text{BZ}}\frac{dk}{2\pi}\partial_k\arg [E_n(k)-E_R],
\end{align}
where $E_n(k)$ represents the eigenenergy of the non-Hermitian Bloch Hamiltonian, $n$ is the band index and the integral runs over the whole Brillouin zone. We obtain $w=\pm1$ in the tight binding limit for the bands having real energy $<0$ and $>0$, respectively. For $w=0$ the system does not show the skin effect and it is topologically trivial. The winding number can be visualized 
by plotting the real energy vs the imaginary energy of the Bloch Hamiltonian $H(k)$. $w\neq0$ corresponds to a loop in the complex energy plane, whereas $w=0$ corresponds to a line or arc. In Fig.~\ref{FigWinding} the real vs imaginary energy is plotted, which shows two loops, having $w=\pm1$. A loop in the complex plane is a signature of the NHSE \cite{PRX.8.031079.2018,J.Phys.Condens.Matter.31.263001.2019}. $w=\pm1$ is associated with a counter clockwise and a clockwise winding, respectively, as we scan the Brillouin zone (see Movie  6). While $w=+1$ corresponds to the localization at the left edge, $w=-1$ corresponds to the localization at the right edge, which is consistent with Figs.~\ref{Fig5}(b) and \ref{Fig5}(c).

We stress that for $\Gamma_{\sigma_+}=\Gamma_{\sigma_-}$, the complex eigenenergies have the same imaginary part. Consequently, they do not form a loop and instead form a straight line parallel to the real energy axis. This also indicates that it is necessary to have  $\Gamma_{\sigma_+}\neq \Gamma_{\sigma_-}$ in order to obtain the NHSE.

\section*{Multi charged vortex states and persistent currents}
\begin{figure*}[t]
\centering
\includegraphics[width=0.7\textwidth]{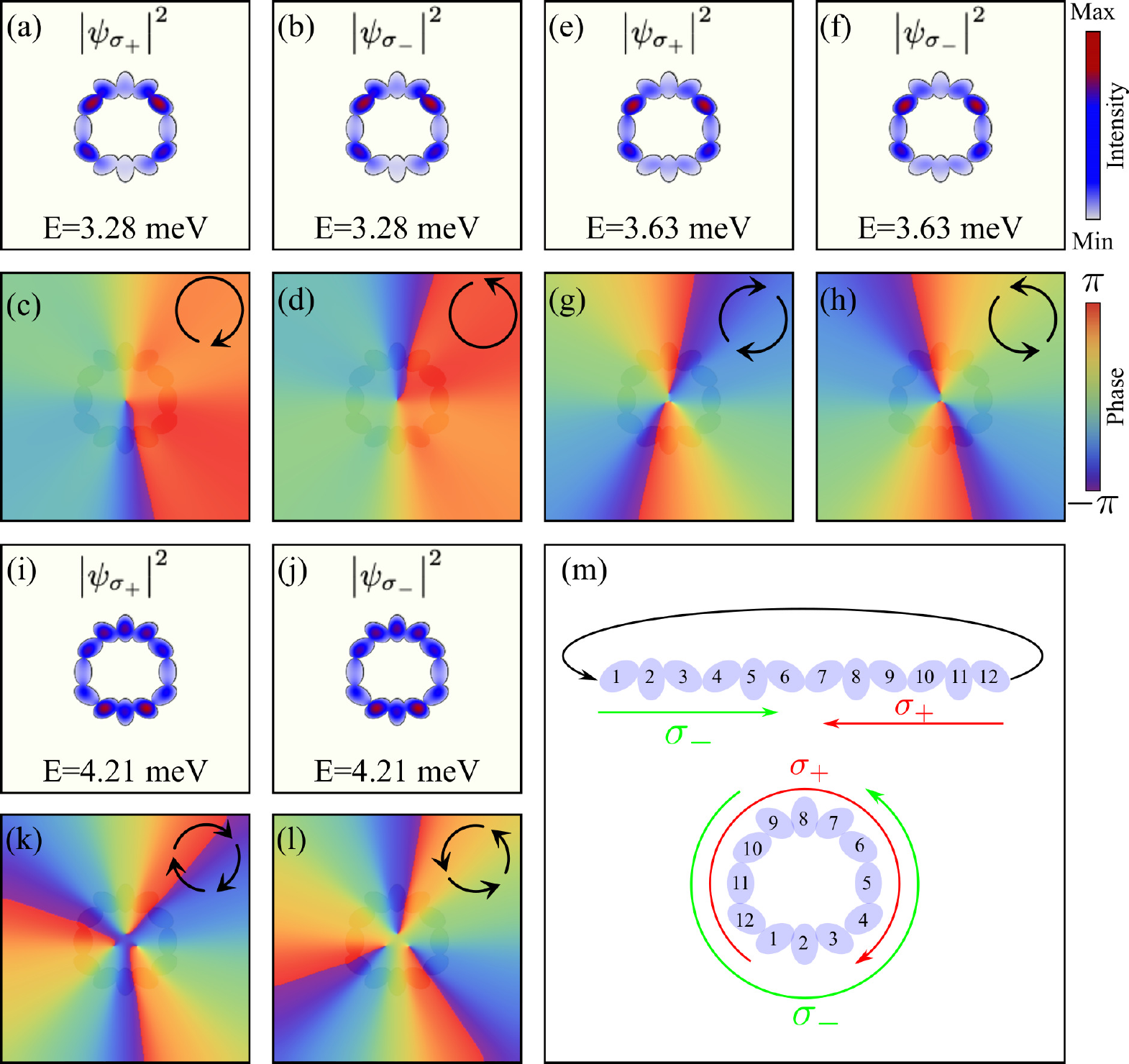}
\caption{ (a)-(l) Vortex and anti-vortex eigenstates corresponding to a circular chain. The vortex states are linked with $\sigma_+$ states, whereas the anti-vortex states are linked with $\sigma_-$ states. The diameter of the circular chain is 4.1 $\mu$m. All other parameters are kept the same as those in Fig.~ \ref{Fig3}. (m) Formation of a circular chain from a linear chain.}
\label{FigVortex}
\end{figure*}

Here instead of a linear chain, we consider a circular chain of elliptical micropillars. Each micropillar is oriented by an angle $\pi/3$ with respect to its previous one. In Fig.~\ref{FigVortex} a circular chain having diameter 4.1 $\mu$m and consisting of 12 micropillars is considered. Due to the non-trivial spin-Hall nature, the eigenstates of the system form vortices having vorticity $\pm1,~\pm2,~\pm3$, etc. The interesting feature of the vortex is that the sign of the vorticity is locked with the circular polarization of the states: all vortices are $\sigma_+$ polarized, while all anti-vortices are $\sigma_-$ polarized. We take charge as positive (negative) for the clockwise (counter-clockwise) rotating vortex. In Fig.~\ref{FigVortex} vortex states having vorticity $\pm1$ (a-d), $\pm2$ (e-h), and $\pm3$ (i-l) are shown. These states have different energies and can be excited using a circularly polarized coherent pump.  In Movie  4, the propagation of the $\sigma_\pm$ polaritons in the clockwise and counter-clockwise directions, respectively, under the effect of a linearly polarized coherent pulse is shown. 

\begin{figure*}[t]
\centering
\includegraphics[width=0.7\textwidth]{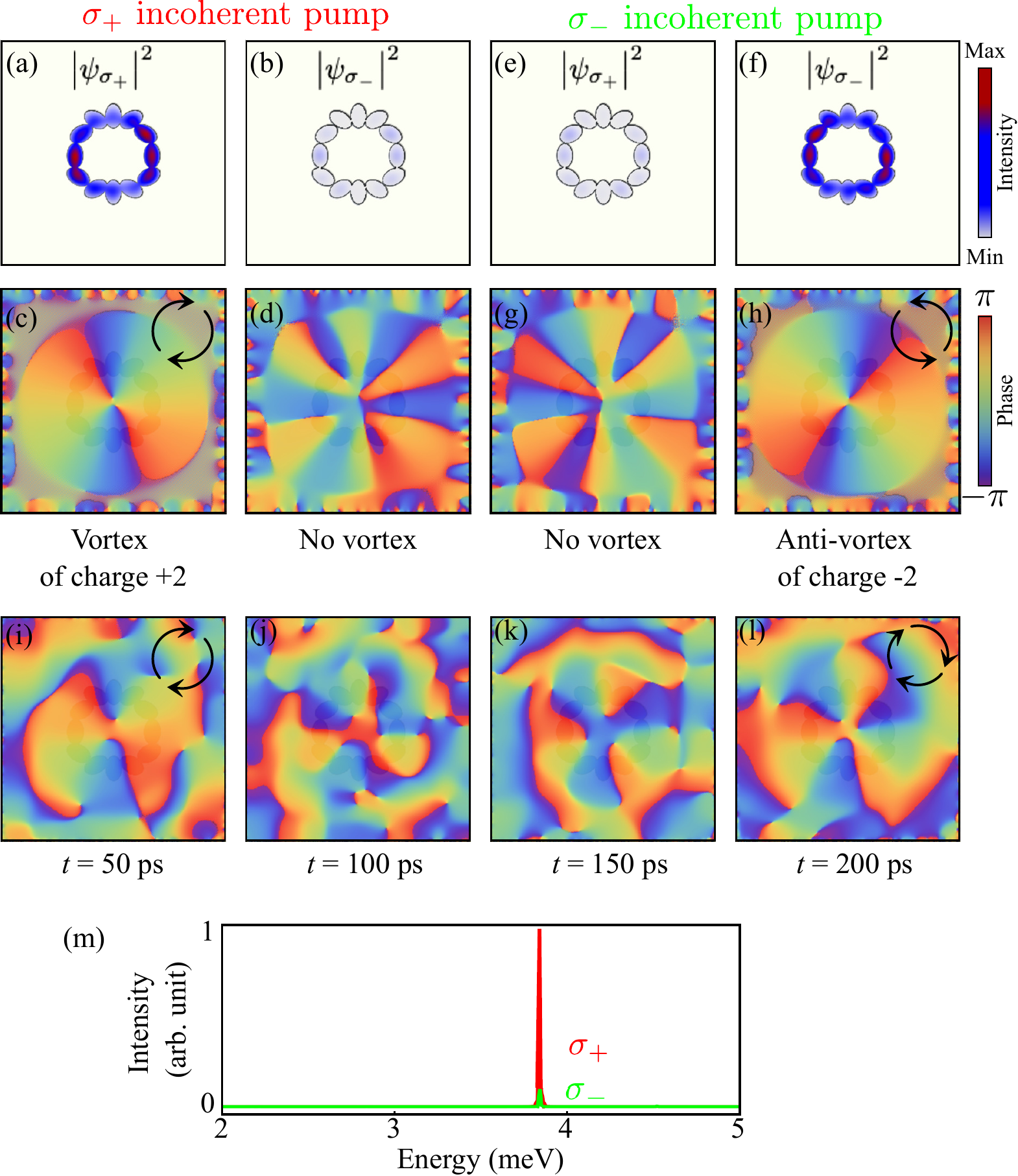}
\caption{ (a)-(h) Spontaneous formation of vortex and anti-vortex states under a circularly polarized incoherent pump. While $\sigma_+$ incoherent pump creates a vortex of charge +2, $\sigma_-$ incoherent pump creates an anti-vortex of opposite charge. (i)-(l) Change of vortex charge by only changing the pump power. (m) Intensity of the polaritons as a function of energy for the vortex formation shown in (a-d). Parameters: (a)-(d) $P_{\sigma_+}=0.08$ ps$^{-1}\mu$m$^{-2}$ and $P_{\sigma_-}=0$. (e)-(h) $P_{\sigma_+}=0$ and $P_{\sigma_-}=0.08$ ps$^{-1}\mu$m$^{-2}$. (i)-(l) $P_{\sigma_+}=0.22$ ps$^{-1}\mu$m$^{-2}$ and $P_{\sigma_-}=0$. All other parameters are kept the same as those in Fig. \ref{Fig5}.}
\label{FigVortex_NonReso}
\end{figure*}

This system is suitable for realization of a vortex source as a polariton condensate forms in one of the vortex states under a circularly polarized incoherent pump. To show this we consider a Gaussian shaped circularly polarized incoherent pump positioned at the center of the chain having full width at the half maximum 7 $\mu$m. The dynamics of the polaritons is given by the following driven dissipative Gross-Pitaevskii equation

\begin{align}
i\hbar\frac{\partial\psi_{\sigma_{\pm}}}{\partial t}&=\left[-(1-i\lambda)\frac{\hbar^2\nabla^2}{2m}+V(x,y)-i\hbar\frac{\gamma}{2}\right]\psi_{\sigma_{\pm}}\notag\\
&+V_T(x,y,\theta)\psi_{\sigma_\mp}+\left(\alpha_1\left|\psi_{\sigma_\pm}\right|^2+\alpha_2 \left|\psi_{\sigma_\mp}\right|^2\right)\psi_{\sigma_\pm}\notag\\
&+\left(g_r+i\hbar\frac{R}{2}\right)n_{\sigma_{\pm}}\psi_{\sigma_{\pm}},\label{SEq5}\\
\frac{\partial n_{\sigma_{\pm}}}{\partial t}&=-\left(\gamma_r+R|\psi_{\sigma_{\pm}}|^2\right)n_{\sigma_{\pm}}+\xi\left(n_{\sigma_\mp}-n_{\sigma_\pm}\right)\notag\\
&+P_{\sigma_{\pm}}\exp\left[-\left[(x-X_0)^2+(y-Y_0)^2\right]/(2\Delta^2)\right]\label{SEq5B}.
\end{align}
Here a phenomenological parameter $\lambda=5.3\times10^{-3}$ is introduced to take into account energy relaxation processes \cite{PRB.80.195309.2009}. This corresponds to higher energy states decaying faster compared to the lower energy states. $(X_0,Y_0)=(0,0)~\mu$m is the position of the incoherent pump and $\Delta=7~\mu$m is its width. We set the pump strength above the condensation threshold and solve Eqs.~(\ref{SEq5}) and (\ref{SEq5B}) using very small random noise as initial condition. Since the sign of the vortex is locked with the polariton spin, we get the spontaneous formation of a vortex of charge +2 under $\sigma_+$ incoherent pump with the condensate being around 85\% $\sigma_+$ polarized [see Figs.~\ref{FigVortex_NonReso}(a), \ref{FigVortex_NonReso}(b), \ref{FigVortex_NonReso}(c), and \ref{FigVortex_NonReso}(d)] and an anti-vortex of charge -2 under $\sigma_-$ incoherent pump with the condensate being around 85\%  $\sigma_-$ polarized  [see Figs.~\ref{FigVortex_NonReso}(e), \ref{FigVortex_NonReso}(f), \ref{FigVortex_NonReso}(g), and \ref{FigVortex_NonReso}(h)]. The formation of the vortex and anti-vortex from the random noise is shown in Movie  5. The charge of the vortex can be controlled by changing the strength and/or $\Delta$ of the incoherent pump. Similar to the NHSE, here also $\sigma_\pm$ polaritons experience different gains under the circularly polarized incoherent pump. However,  since the polaritons do not experience any boundary along their propagation direction, unlike in the NHSE here they do not show any localization effect.  In contrast to previous theoretical work on the generation of multiply charged vortices \cite{PRB.88.201303.2013,Optica.8.301.2021}, here condensation into a multiply charged vortex state of given sign would be attained deterministically. The result could be interpreted also as a persistent current in the system, but again with a deterministic rather than stochastically chosen direction \cite{PRL.119.067406.2017}.

By increasing the incoherent pump power ($P_{\sigma_+}$) the charge of the vortex can be changed to 3  without changing any other parameter in our model. This clearly is a non-linear effect as the increase in the pump power increases the intensity of the polaritons and  is illustrated in Fig.~\ref{FigVortex_NonReso}(i)-(l), where at initial times the system shows a signature of a vortex of charge +2. However, for longer times when the polariton intensity increases the system shows a steady state of a vortex of charge +3. We stress that such a change in the vortex charge is not possible by only changing the pump power in a linear system. This feature also makes the present system optically tuneable for creating vortices of different charges. The non-linear nature of the polaritons also makes the system suitable for studying the coupling between differently charged vortices under a coherent pump \cite{OptLett.45.5700.2020}.

In our calculations, in writing the driven dissipative Gross-Pitaevskii equation, we have assumed that polaritons form in a single coherent state such that the mean-field approximation is valid, as is common in the description of polariton condensates \cite{RevModPhys.85.299.2013}. In reality, polariton condensates have imperfect coherence, as multiple modes may be simultaneously populated such that regular polariton condensates have a second order correlation function, $g_2\approx1.1$ \cite{PRL.101.067404.2008}. If special effort is made to isolate a single mode for condensation or polariton lasing one can attain \cite{PhysRevX.6.011026.2016} near perfect coherence $g_2-1=10^{-3}$  and indeed in lattices such as Hermitian SSH chains (where topologically protected modes are at the edges of the system and somewhat separated from others) good coherence is also attained \cite{ACS.Photon.8.1377.2021}. We note however, that such isolation of modes does not occur in the non-Hermitian skin effect where they all remain overlapping spatially. We thus anticipate that the multi-charged vortex states would have the same coherence of regular polariton condensates. However, there is no work, to our knowledge that has been able to isolate a single vortex mode in a polariton condensate, without the possibility of other overlapping, albeit weakly populated, modes. In principle, coherence can be calculated with density matrix approaches \cite{ACS.Photon.8.1377.2021}, however, for systems where many modes are simultaneously overlapping Monte Carlo \cite{PhysRevB.92.115117.2015} or Bogoliubov methods \cite{PhysRevX.10.041060.2020} might be more appropriate.

We can attain the energy dependent intensity of the polaritons by Fourier transforming the wave function $\psi_{\sigma_\pm}$ from Eq.~(\ref{SEq5}) along the time axis and integrating the intensity along the spatial axes.  Fig.~\ref{FigVortex_NonReso}(m) shows the intensity distribution as a function of energy for the vortex state shown in Fig. \ref{FigVortex_NonReso}(a)-(d). The intensity has a peak around energy 3.63 meV, which is a vortex state having charge 2 [ Fig. \ref{FigVortex}(e)-(h)], showing the single-mode behaviour of the system. The presence of a single mode condensate is promising for coherence, although, as we have mentioned, other theoretical techniques would be needed to access coherence.

\section*{Effect of the degree of circular polarization of the excitonic reservoir}
In this section, we discuss the effect of the degree of circular polarization of the reservoir on the localization of the skin modes. In order to obtain the localization profile of the skin modes we evolve Eqs.~(\ref{Eq4}) and (\ref{Eq4B}) for $P_{\sigma_+}\neq0$ and $P_{\sigma_-}=0$. After the steady state of the reservoir ($\partial n^s_{\sigma_\pm}/\partial t$=0) is reached we diagonalize the Hamiltonian corresponding to the following equation

\begin{align}
&\left[-\frac{\hbar^2\nabla^2}{2m}+V(x,y)-i\hbar\frac{\gamma}{2}\right]\psi^E_{\sigma_{\pm}}+V_T(x,y,\theta)\psi^E_{\sigma_\mp}\notag\\
&~~~~~~~~+\left(g_r+i\hbar\frac{R}{2}\right)n^s_{\sigma_{\pm}}\psi^E_{\sigma_{\pm}}=E\psi^E_{\sigma_\pm}.
\end{align}
Here $\psi^E_{\sigma_\pm}$ is the eigenstate having eigenenergy $E$ and $n^s_{\sigma_\pm}$ is the steady state density of the reservoir. The degree of the circular polarization of the reservoir is given as 
\begin{align}
S_z^{\text{Res}}(\%)=\frac{n^s_{\sigma_+}-n^s_{\sigma_-}}{n^s_{\sigma_+}+n^s_{\sigma_-}}\times 100.
\end{align}

\begin{figure*}[t]
\centering
\includegraphics[width=0.8\textwidth]{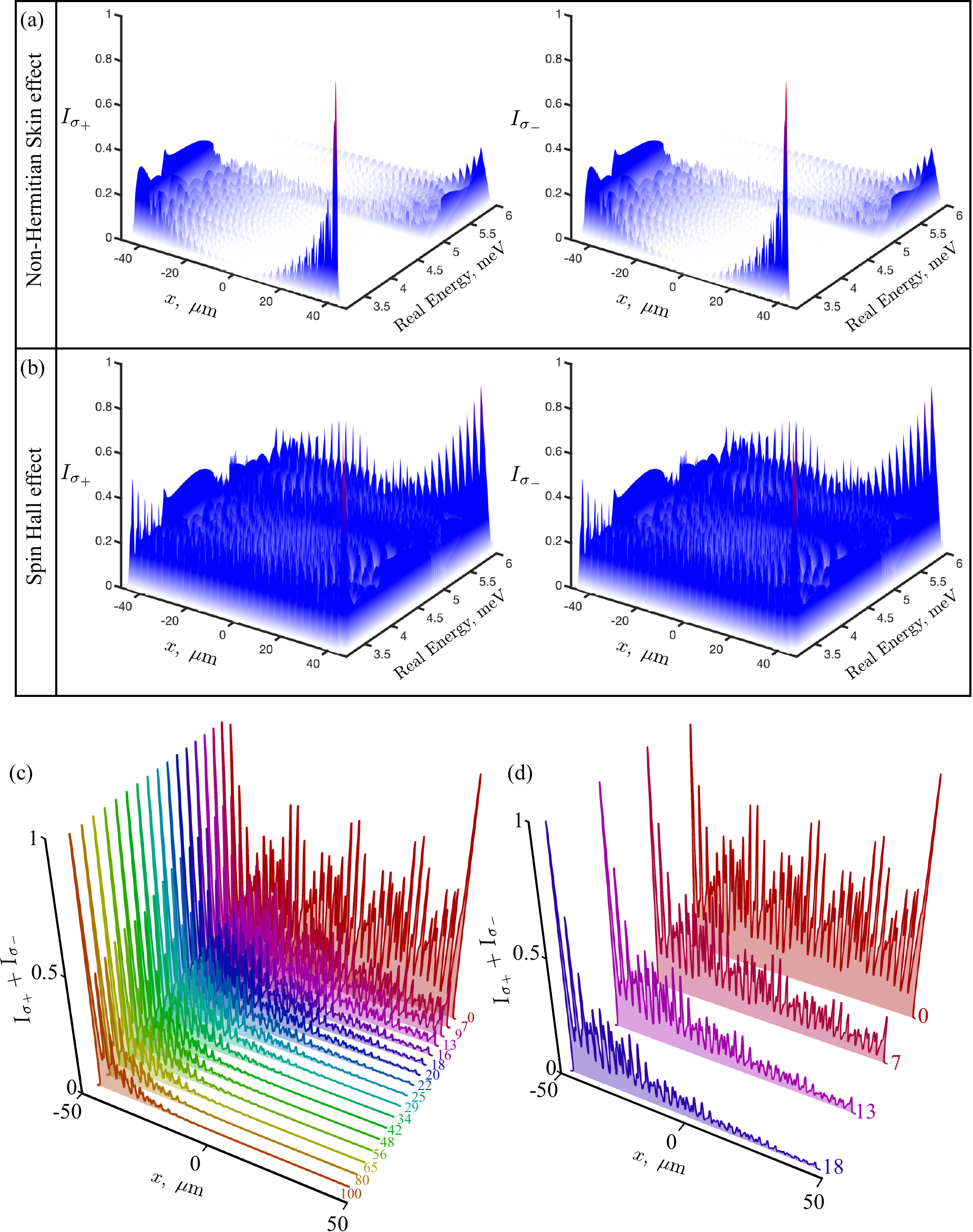}
\caption{(a)-(b) Localization of the modes in the NHSE ( $S_z^\text{Res}=18\%$) and TSHE ($0\%$), respectively. (c) Localization of the mode having energy 4 meV for different values of $S_z^{\text{Res}}$. $\xi=\left[0, 0.1, 0.2,0.3,0.4,0.5,0.7,0.9,1.1,1.3,1.5,1.7,1.9,2.5,3.5,4.5,20\right]$ ps$^{-1}$ correspond to $S_z^{\text{Res}}=\left[100\%,80\%,65\%,56\%,48\%,42\%,34\%,29\%,25\%,22\%,20\%,18\%,16\%,13\%,9\%,7\%,0\%\right]$, respectively. (d) Zoomed view of the localization profiles in (c) for $S_z^{\text{Res}}=18\%, 13\%, 7\%~\text{and}~ 0\%$.}
\label{FigLocalizationReservoir}
\end{figure*}

It should be noted that due to the presence of the spin relaxation term ($\xi$ in Eq.~(\ref{Eq4B})) in the reservoir dynamics, $n^s_{\sigma_-}\neq 0$ although $P_{\sigma_-}=0$. If the reservoir has non-zero $S_z^{\text{Res}}$, the gain corresponding to $\sigma_\pm$ polaritons are different $\left(i\hbar Rn^s_{\sigma_{+}}/2\neq i\hbar Rn^s_{\sigma_{-}}/2\right)$, which is essential for the NHSE.

In Figs.~\ref{FigLocalizationReservoir}(a) and \ref{FigLocalizationReservoir}(b) the i ntensity profiles of the modes $I_{\sigma_\pm}(x)=\int|\psi^E_{\sigma_\pm}|^2~dy$ are plotted as a function of $x$ and $E$ for $S_z^{\text{Res}}\approx18\%$ and $S_z^{\text{Res}}\approx0\%$, respectively. The plot in Fig.~\ref{FigLocalizationReservoir}(a) is the same as Figs.~\ref{Fig5}(d) and \ref{Fig5}(e), but with a three dimensional view. The modes are delocalized for $S_z^{\text{Res}}\approx0\%$, which is an indication of the absence of skin effect. This case corresponds to the TSHE case, where the non-Hermitian gain and decay terms play the role to increase and decrease the intensity of the polariton with time, respectively.

The value of $S_z^{\text{Res}}$ is difficult to control in experiments as it is determined by many factors, such as working temperature, sample material, detuning of the circularly polarized incoherent laser with the exciton energy, etc. We note that the preservation of the spin under non-resonant pumping was shown first in planar microcavities under both pulsed \cite{PRL.109.016404.2012} and continuous wave excitation \cite{PRL.109.036404.2012}. Indeed, though the preservation of the spin was not perfect and there was some spin relaxation. It has been shown, more recently, in micropillar samples (such as the ones considered in our work), that 100\% circular polarization of a polariton condensate is achievable with circularly polarized non-resonant pumping \cite{PRB.99.115303.2019}. Still, we need to be careful that this does not prove that the reservoir is fully spin polarized. Although a recent experiment has shown this value to be around 18$\%$ \cite{NatPhoton.13.283.2019}, here we study the effect of $S_z^{\text{Res}}$ on the mode localization. By varying $\xi$ in Eq.~(\ref{Eq4B}) we can adjust the value of $S_z^{\text{Res}}$ numerically. In Fig.~\ref{FigLocalizationReservoir}(c) the mode profiles of the mode having energy 4 meV are plotted for different values of $S_z^{\text{Res}}$. It shows that the mode is very well localized for $S_z^{\text{Res}}=100\%$ and starts to delocalize as  $S_z^{\text{Res}}$ decreases. A zoomed view of the mode localizations corresponding to 18\%, 13\%, 7\% and 0\% of the $S_z^{\text{Res}}$ values are shown in Fig.~\ref{FigLocalizationReservoir}(d). It shows that the localization is weak, but it still exists even for $S_z^{\text{Res}}=7\%$. As expected the mode is delocalized for $S_z^{\text{Res}}=0\%$. Since the signature of the skin effect remains for partially polarized reservoir, even for very small $S_z^{\text{Res}}$, we expect this can be measured in experiments.  Alternatively, the localization length can also provide an estimation of $S_z^{\text{Res}}$, which is difficult to access in experiments otherwise.

\section*{Discussion on the non-reciprocal nature of the NHSE}
\begin{figure*}[t]
\centering
\includegraphics[width=0.8\textwidth]{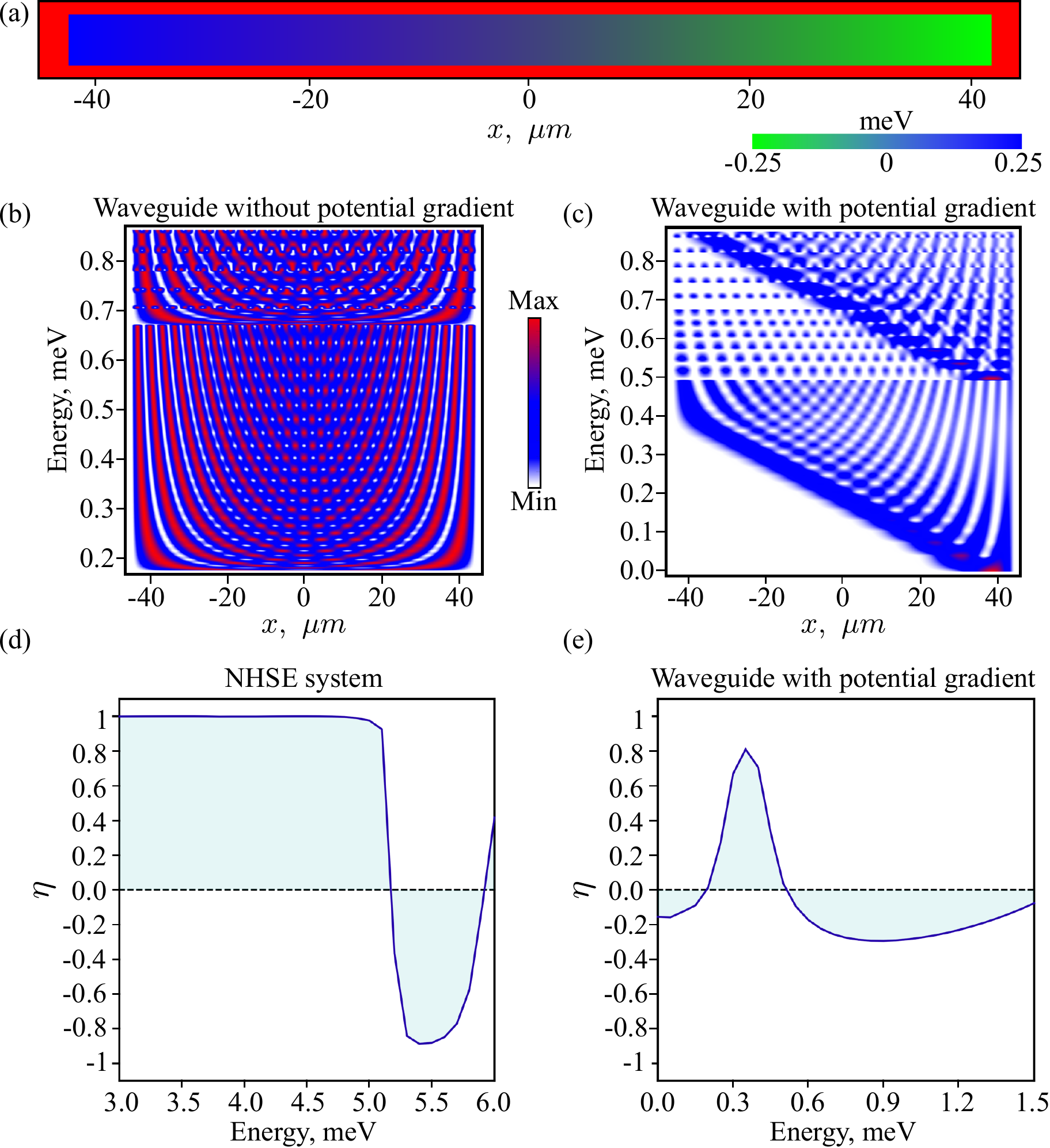}
\caption{(a) A waveguide having 0.5 meV potential gradient along the $x$ axis. The red region corresponds to a 10 meV potential. (b)-(c) Spatial profiles of the modes of the waveguide without and with the potential gradient, respectively. (d)-(e) The degree of non-reciprocity for  NHSE and potential gradient waveguide system, respectively. The NHSE shows very large non-reciprocity over a wide range of energy. The coherent pump parameters are the same as those used in Fig.~\ref{Fig4}.}
\label{FigNonReciporcity}
\end{figure*}

Here we quantify and compare the non-reciprocal nature of the NHSE. For comparison we choose a polariton waveguide having a potential gradient of 0.5 meV over its whole length. In Fig.~\ref{FigNonReciporcity}(a) such a waveguide is shown. The total system size of the waveguide is kept the same as the one considered for elliptical pillar chain. The red region corresponds to a 10 meV potential, similar to the elliptical pillars. In Figs.~\ref{FigNonReciporcity}(b) and \ref{FigNonReciporcity}(c) the spatial profiles of the modes of the waveguide are shown without and with the potential gradient. Without the potential gradient, all the modes of the waveguides are delocalized into the bulk, while with the potential gradient lower energy modes are indeed localized at the right end of the waveguide, which is the lower potential region. However, at higher energies the modes become delocalized into the bulk.

Next to  quantify the non-reciprocity, we perform a set of numerical experiments. We consider a continuous coherent pump placed at the right end of the elliptical micropillar chain (similar to Figs.~\ref{Fig5}(j) and \ref{Fig5}(k)) and record the intensity $I_{R\rightarrow L}(\hbar\omega)$ at the left end of the chain. We scan the whole band in Fig.~3(a) by changing the energy $\hbar\omega$ of the coherent pump. Next we place the same coherent pump at the left end of the chain (similar to Figs.~\ref{Fig5}(f) and \ref{Fig5}(g)) and record the intensity $I_{L\rightarrow R}(\hbar\omega)$ at the right end of the chain. We once again scan the whole band by changing the energy $\hbar\omega$ of the pump. The expression for $I_{R\rightarrow L}$  can be expressed as

\begin{align}
I_{R\rightarrow L}=\int |\psi_{\sigma_\pm}(x,y)|^2\times e^{-\frac{(x-X_0)^2+(y-Y_0)^2}{(2\Delta^2)}}~dxdy.\label{EqS8}
\end{align}
Here $\psi_{\sigma_\pm}(x,y)$ is the steady state polariton  wave function under the coherent pump, $(X_0,Y_0)=(-45,0)~\mu$m is the mean position of the Gaussian and $\Delta=10~\mu$m is the width of the Gaussian. The expression for $I_{L\rightarrow R}$ remains the same as the one in Eq.~(\ref{EqS8}), but with the position of the Gaussian $(X_0,Y_0)=(45,0)~\mu$m, which is at the right end of the chain. The degree of the non-reciprocity $\eta$ is given by

\begin{align}
\eta=\frac{I_{R\rightarrow L}-I_{L\rightarrow R}}{I_{R\rightarrow L}+I_{L\rightarrow R}}.
\end{align}

For a reciprocal system $I_{R\rightarrow L}=I_{L\rightarrow R}$ and consequently $\eta=0$. For a non-reciprocal system $\eta\neq 0$. In Fig.~\ref{FigNonReciporcity}(d) $\eta$ is plotted as a function of energy for the NHSE. The system shows perfect non-reciprocity $\eta\approx 1$ for a wide range of energy. $\eta\rightarrow -1$ for the higher energies, which is understandable from the localization of the modes shown in Fig.~\ref{FigLocalizationReservoir}(a). 

Next we follow the same procedure to obtain $\eta$ for the waveguide with a potential gradient. In Fig.~\ref{FigNonReciporcity}(e) $\eta$ is plotted as a function of energy. Due to the potential gradient the lower energy modes are located at the right hand side, making $\eta$ negative and as energy increases, the modes start to become delocalized and get located at the left end, making $\eta$ positive. Near 0.5 meV the $p$ modes start, which are located at the right end along with the delocalized $s$ modes. As a result $\eta$ decreases and becomes negative. The maximum value of $\eta$ is around 0.8, however the energy window showing such a value is very small. The reason that $|\eta|$ does not approach 1 around energies where the wave function is localized at the right end is because those modes are near the ground state modes. As a result they do not have significant group velocity for the polaritons to travel from one end to other end. At relatively higher energies,  the group velocity increases, but the wave functions also become delocalized. On the other hand, for the NHSE the modes that are localized are effectively the edge states of a two dimensional topological system, which have large group velocity and ideal for propagation. As a result $\eta\approx 1$ for a wide range of energy, making it more robust and efficient for transporting polariton in one way compared to the waveguide system. One other advantage of the NHSE is that the lattice can in principle be infinitely long. However, there is a limit to the length of the waveguide structure with potential gradient as the variation of the photon potential can not be more than a fraction of the Rabi splitting without changing too much the photonic and exciton fractions and potentially losing the existence of polaritons.

\section*{Conclusion} 
We have presented a scheme based on an elliptical micropillar chain, where the naturally present polarization splitting inside the micropillars along with their orientation give rise to the topological spin Hall effect, where $\sigma_\pm$ polaritons propagate in opposite directions. Being one dimensional, the presented scheme is more compact than their two dimensional counterpart \cite{Science.370.600.2020,PRB.103.L201406.2021}. In the presence of a circularly polarized uniform incoherent pump, the non-Hermitian physics takes over and the topology changes from the spin-Hall effect to a non-Hermitian skin effect (NHSE), where the polaritons accumulate at one end of the chain independent of the excitation position.  A recent paper \cite{PRL.125.123902.2020} proposed the NHSE using tightly localized alternating $\sigma_\pm$ incoherent pumps subjected at two elliptical micropillars, where within one micropillar pair the effective coupling is only one way (non-reciprocal). By joining such micropillar pairs to make a chain, the Hatano and Nelson model \cite{PRL.77.570.1996} can be realized, which leads to the NHSE. Although the scheme is unique and theoretically interesting, its experimental realization may be challenging. We believe that our present scheme is much more simple and realistic for experiments, which relies upon a uniform circularly polarized incoherent pump (instead of  tightly localized alternating $\sigma_\pm$ pumps) to obtain the NHSE.

As a combination of the above two effects, if a circular instead of a linear chain is considered, vortex-antivortex states having different charges arise. The special feature in this system is that while all the vortex states are locked to $\sigma_+$ spin, all antivortex states are locked with $\sigma_-$ spin. This gives the advantage of controlling the sign of the vortex using the spins. We further demonstrate that it is possible to form polariton condensates in these states using a circularly polarized incoherent pump, effectively using it as a vortex source.  

The circular polarization of the excitonic reservoir plays an  important role, but does not need to be 100\%. In fact, by looking at the localization of the states one could extract the reservoir polarization, which has not been measured directly in experiment. Given that the realization of the all optical polariton lattices \cite{Communications.Phys.3.2.2020,PRL.124.207402.2020,PRB.103.155302.2021} and topological phases \cite{NatCom.11.4431.2020,Optica.8.1084.2021} have started to become popular in recent times, our work lays a perfect platform for extending the field in the non-Hermitian regime.  Apart from the fundamental interest, the non-trivial directional dependent motion of the spins in the TSHE, may find its application in connecting different spin-dependent polariton devices \cite{NatPhoton.4.361.2010,APL.107.011106.2015}. On the other hand, the large non-reciprocity in the NHSE can be used to suppress feedback in polariton networks \cite{PRAppl.11.064029.2019,NanoLett.20.3506.2020,PRAppl.13.064074.2020,Nano.Lett.Unnamed.2021} and connecting different components of a polariton information processing device \cite{NatCom.4.1778.2013,APL.101.261116.2012,APL.103.201105.2013,NatCom.11.897.2020}.

\section*{acknowledgement}
We thank Christian Schneider (Universit\"{a}t Oldenburg) and Sebastian Klembt (University of W\"{u}rzburg) for stimulating discussions. 

\section*{Funding Sources}
The work was supported by the Ministry of Education, Singapore (Grant No. MOE2019-T2-1-004).

\end{document}



\begin{abstract}
In this supporting information we provide some additional calculations such as the derivation of the realization of the complex hopping phase, polariton condensation in a linear chain, and polariton dynamics under a coherent pulsed excitation.
\end{abstract}

\section{Realization of the complex hopping between $\sigma_\pm$ modes}
To show the complex hopping between the $\sigma_\pm$ spin states inside an elliptical micropillar, we consider $L$ and $T$ to be the energies of the modes having linear polarization along the longitudinal (semi-major axis) and transverse (semi-minor axis) directions, respectively. This can be represented as 
\begin{align}\label{SEq1}
i\hbar\frac{\partial}{\partial t}\begin{bmatrix}\psi_L\\
 \psi_T\end{bmatrix}=\begin{bmatrix}L &0\\
 0&T\end{bmatrix}\begin{bmatrix}\psi_L\\
 \psi_T\end{bmatrix}=H_{LT}\begin{bmatrix}\psi_L\\
 \psi_T\end{bmatrix},
\end{align}
where
\begin{align}\label{SEq2}
H_{LT}=\begin{bmatrix}L &0\\
 0&T\end{bmatrix}.
\end{align}

If the longitudinal axis of the elliptical micropillar makes an angle $\theta$ with the $x$ axis, then in the circular polarization basis the system can be represented as
 \begin{align}\label{SEq3}
i\hbar\frac{\partial}{\partial t}\begin{bmatrix}\psi_{\sigma_+}\\
 \psi_{\sigma_-}\end{bmatrix}&=M^{\dagger}H_{LT}M\begin{bmatrix}\psi_{\sigma_+}\\
 \psi_{\sigma_-}\end{bmatrix}=\begin{bmatrix}\varepsilon &\Delta e^{2i\theta}\\
 \Delta e^{-2i\theta}&\varepsilon\end{bmatrix}\begin{bmatrix}\psi_{\sigma_+}\\
 \psi_{\sigma_-}\end{bmatrix}.
\end{align}
Here
\begin{align}\label{SEq4}
M=\frac{1}{\sqrt{2}}\begin{bmatrix}e^{-i\theta} &e^{i\theta}\\
 -ie^{-i\theta} &ie^{i\theta}\end{bmatrix}
\end{align}
is the transformation matrix, $\varepsilon=\left(L+T\right)/2$ represents the onsite energy, and $\Delta=\left(L-T\right)/2$ represents the polarization splitting. It shows that $\sigma_\pm$ spin states are coupled with a complex hopping, where an orientation angle  $\theta$ induces $2\theta$ phase in the hopping term.

\section{Exciton-polariton condensation}
\begin{figure}[H]
\includegraphics[width=\textwidth]{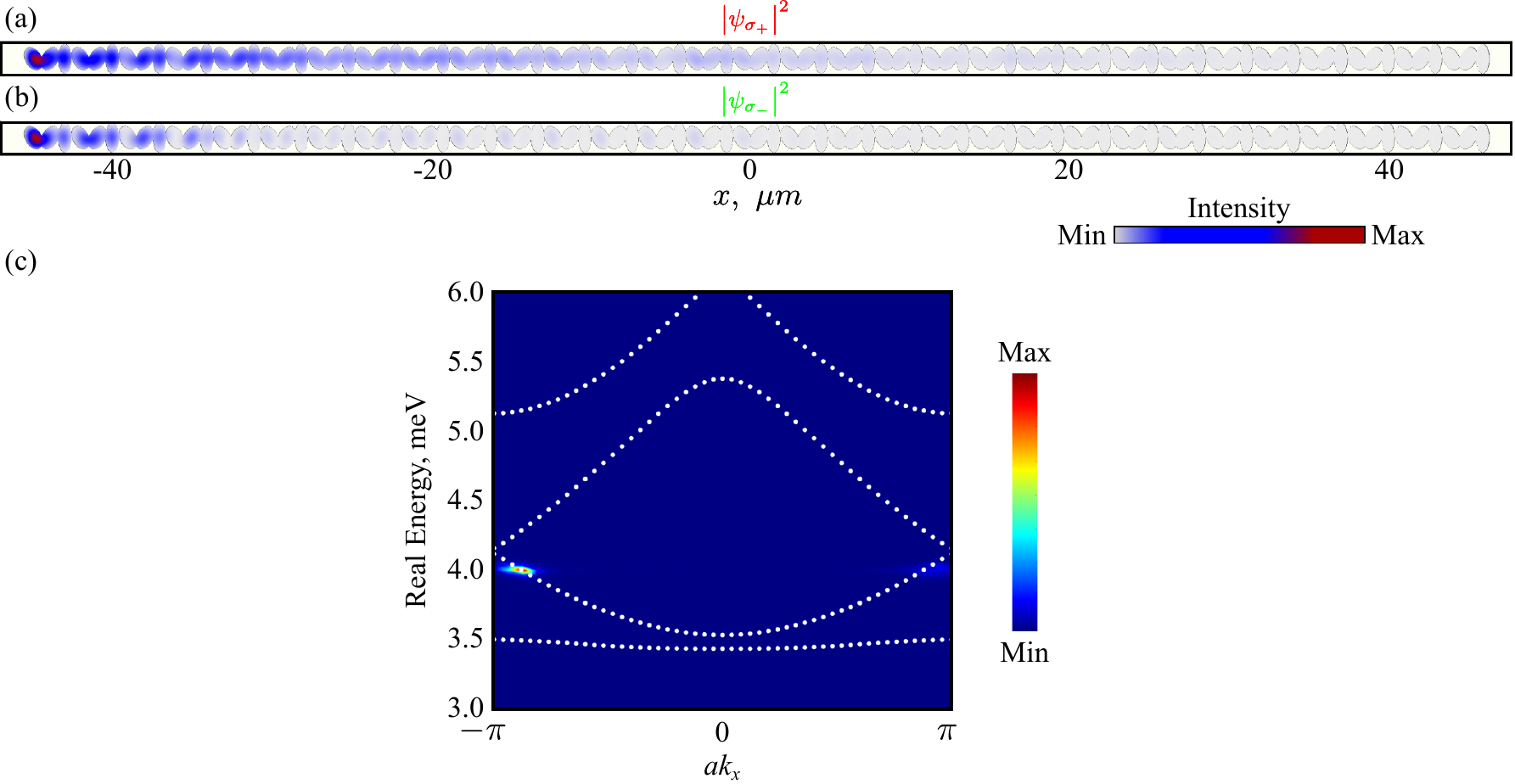}
\caption{(a)-(b) Intensities of the condensate corresponding to the $\sigma_\pm$ modes. (c) Formation of the condensate in the reciprocal space. The dotted curve is the dispersion corresponding to the linear case shown in Fig. 3 in the main text. All the parameters, except the incoherent pump being spatially uniform, are kept the same as those in Fig.~8 in the main text.}
\label{SuppF1}
\end{figure}

In order to show the exciton-polariton condensation, we set the spatially uniform incoherent pumps $P_{\sigma_+}=0.08~\text{ps}^{-1}\mu\text{m}^{-2}$ and $P_{\sigma_-}=0$, which is above the condensation threshold.  We keep all other parameters the same as those in Figs.~5(f)-(k) of the main text and solve the time dynamics in Eqs.~(6)-(7) in the main text using a very small random noise as an initial condition. In Figs.~\ref{SuppF1}(a-b) the intensities of the $\sigma_\pm$ modes are shown, respectively. Although $P_{\sigma_+}$ is uniformly distributed over the whole lattice, the condensates form at the left end of the lattice due to the non-Hermitian skin effect. This is in contrast to the vortex formation for a circular chain in Fig.~8 of the main text, where due to the absence of boundary no localization is observed. In Fig.~\ref{SuppF1}(c) the dispersion corresponding to the condensate is shown, which is slightly blueshifted due to the polariton-polariton interaction. Similar topological edge mode condensation, but based on Hermitian Su-Schrieffer-Heeger polariton lattice has been demonstrated recently \cite{ACS.Photon.8.1377.2021}.

\section{Comparison of the polariton propagation under a pulsed excitation}
In this section, we compare the dynamics of the polaritons under a pulsed excitation in the topological spin-Hall case and non-Hermitian skin effect case. The dynamics of the polaritons is given by  Eq.~(2-3) in the main text, however the coherent continuous pump term is changed to a pulse $F_{\sigma_\pm}(x,y) e^{\left[-\left(t-t_0\right)^2/2\tau^2\right]}e^{i\left(k_p x-\omega_p t\right)}$.

\begin{figure}[H]
\includegraphics[width=\textwidth]{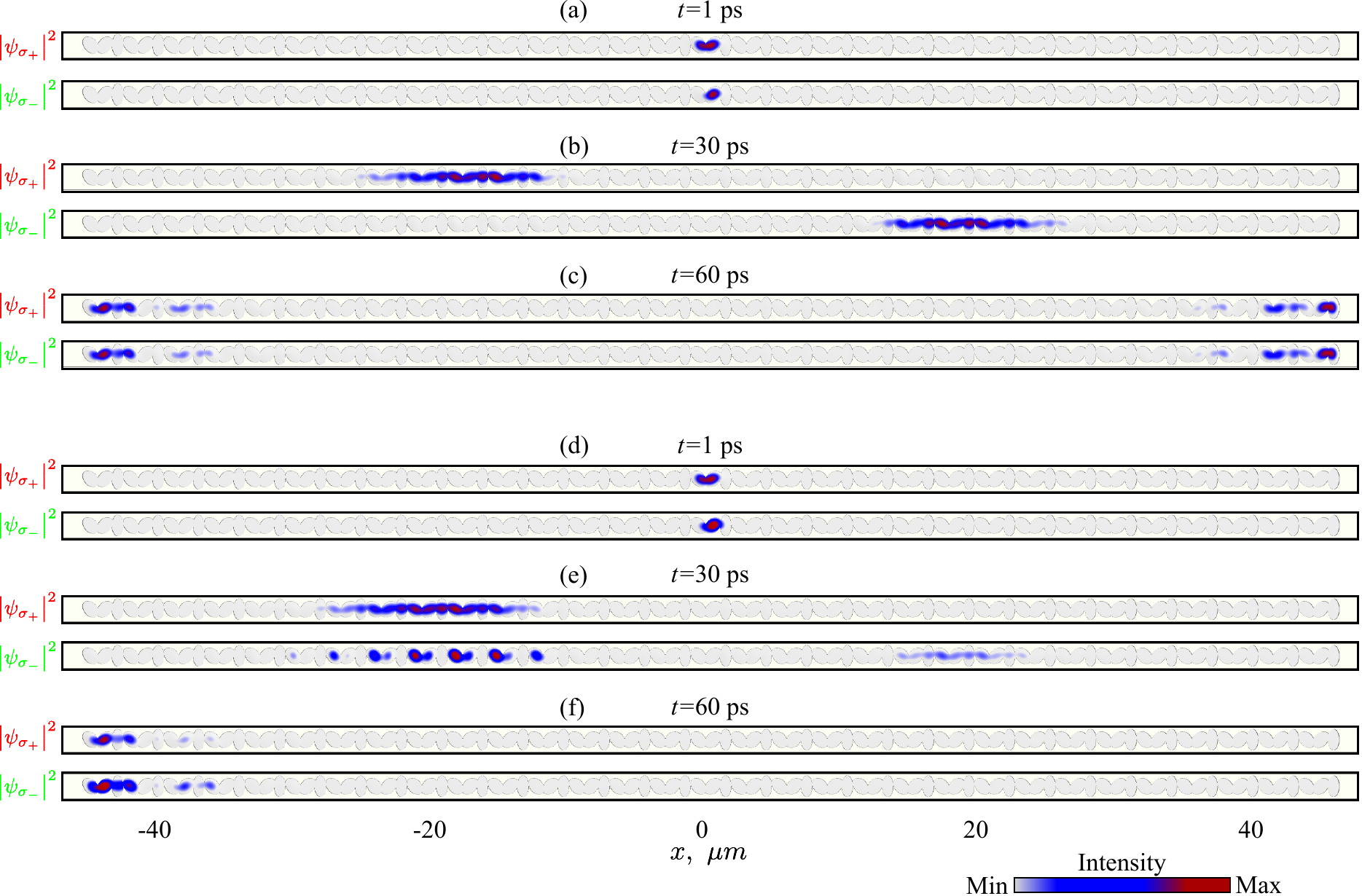}
\caption{Polariton propagation under a pulsed excitation showing the topological spin-Hall effect in (a)-(c) and non-Hermitian skin effect in (d)-(f). Parameters: (a)-(c) all the parameters are the same as those in Fig. 4 of the main text. (d)-(f) all the parameters are the same as those in Fig. 5(f)-5(k) of the main text.}
\label{SuppF2}
\end{figure}

Here $t_0=0$ ps is the time when the pulse is launched and $\tau=10$ ps is the width of the pulse in time. In Fig.~\ref{SuppF2} the dynamics of the polaritons is shown. In Figs.~\ref{SuppF2}(a-c) the spin-Hall nature ($g_r=R=\gamma_r=\xi=P_{\sigma_\pm}=0$) can be seen, where $\sigma_\pm$ polaritons propagate in the opposite directions. When the $\sigma_+$ ($\sigma_-$) polaritons hit the left (right) end of the chain, they couple with the $\sigma_-$ ($\sigma_+$) polaritons and propagate towards the right (left) end. The whole dynamics is shown in movie1 lower panel.

Next in Figs.~\ref{SuppF2}(d-f) the polariton dynamics in the non-Hermitian skin effect regime $\left(P_{\sigma_+}\neq 0,~P_{\sigma_-}=0\right)$ is shown. In this case, $\sigma_\pm$ polaritons propagate in the opposite directions like the TSHE case. However, $\sigma_-$ modes decay very rapidly. As a result after long time all the population is located at the left end of the chain. It should be noted that most of the $\sigma_-$ population at the left end appears due to the spin flip process of the $\sigma_+$ modes. 
 
